\documentclass[reprint,amsmath,amssymb,aps,pra]{revtex4-2}

\usepackage{graphicx}
\usepackage{dcolumn}
\usepackage{bm}

\usepackage{tikz}
\usetikzlibrary{arrows.meta, positioning, calc, shapes.geometric}

\tikzstyle{block} = [draw, fill=blue!10, rectangle, minimum height=2em, minimum width=3em]
\tikzstyle{input} = [coordinate]
\tikzstyle{output} = [coordinate]

\usepackage{algorithm}
\usepackage{algpseudocode}
\newtheorem{remark}{Remark}

\begin{document}

\preprint{APS/123-QED}

\title{Robust Variational Ground-State Solvers via Dissipative Quantum Feedback Models}

\author{Yunyan Lee}
\email{Yun-Yan.Lee@anu.edu.au}
\affiliation{School of Engineering, The Australian National University, Canberra, ACT 2601, Australia}

\author{Ian R. Petersen}
\email{ian.petersen@anu.edu.au}
\affiliation{School of Engineering, The Australian National University, Canberra, ACT 2601, Australia}

\author{Daoyi Dong}
\email{daoyidong@gmail.com}
\affiliation{Australian Artificial Intelligence Institute, Faculty of Engineering and Information Technology, University of Technology Sydney, NSW 2007, Australia}

\date{\today}

\begin{abstract}
We propose a variational framework for solving ground-state problems of open quantum systems governed by quantum stochastic differential equations (QSDEs). This formulation naturally accommodates bosonic operators, as commonly encountered in quantum chemistry and quantum optics. By parameterizing a dissipative quantum optical system, we minimize its steady-state energy to approximate the ground state of a target Hamiltonian. The system converges to a unique steady state regardless of its initial condition, and the design inherently guarantees physical realizability. To enhance robustness against persistent disturbances, we incorporate $H^\infty$ control into the system architecture. Numerical comparisons with the quantum approximate optimization algorithm (QAOA) highlight the method’s structural advantages, stability, and physical implementability. This framework is compatible with experimental platforms such as cavity quantum electrodynamics (QED) and photonic crystal circuits.
\end{abstract}

\maketitle

\section{Introduction}
Optimization problems are central to science and engineering, underlying a wide range of tasks in machine learning, control, and physical modeling~\cite{boyd2004convex, goodfellow2016deep}. In recent years, quantum algorithms have been developed to accelerate specific classes of optimization problems, particularly those involving high-dimensional or combinatorial structures~\cite{biamonte2017quantum, farhi2014quantum, brandao2017quantum}. Among them, identifying the ground state of a quantum Hamiltonian represents an important subclass of quantum optimization problems, with broad applications in quantum chemistry, condensed matter physics, and quantum information science~\cite{yung2014introduction, helgaker2013molecular, bauer2020quantum, mcardle2020quantum, georgescu2014quantum, preskill2018quantum}. Ground-state information is essential for predicting molecular stability, electronic structure, and reaction dynamics, and also serves as a key objective in quantum simulation and variational design~\cite{reiher2017elucidating, cao2019quantum}.

Several quantum algorithms have been proposed to compute ground states of quantum Hamiltonians, including quantum phase estimation (QPE)~\cite{cleve1998quantum}, adiabatic state preparation (ASP)~\cite{farhi2000quantum, albash2018adiabatic, aspuru2005simulated}, and the variational quantum eigensolver (VQE)~\cite{peruzzo2014variational, mcclean2016theory, cerezo2021variational, kandala2017hardware,wang2024entanglement}. Among these, VQE is widely studied due to its suitability for noisy intermediate-scale quantum (NISQ) devices~\cite{tilly2022variational, bharti2022noisy}. Nonetheless, its performance is often limited by hardware noise, circuit depth, and sensitivity to initial-state preparation
~\cite{wang2021noise, arrasmith2021effect, cerezo2021cost, larocca2022diagnosing}. These factors pose challenges for scalability, particularly in analog platforms with limited coherence and restricted control precision~\cite{endo2021hybrid}.

Moreover, VQE and related approaches are typically formulated in terms of qubit-based circuit models. However, many physical systems—particularly in quantum chemistry and quantum optics—are more naturally described using bosonic operators and second quantization~\cite{mcardle2020quantum}. Mapping such systems to spin models can obscure their underlying structure and introduce additional approximation errors~\cite{tilly2022variational, moll2018quantum, cao2019quantum}. These limitations are especially relevant in quantum optical systems, where accurate state initialization and control of continuous-variable modes remain experimentally demanding~\cite{lvovsky2009continuous, weedbrook2012gaussian}. Methods that directly incorporate bosonic degrees of freedom are therefore better aligned with the physics of these platforms and offer improved feasibility for implementation.

In this work, we present a variational framework based on quantum stochastic differential equations (QSDEs)~\cite{zhang2022linear, gardiner2004quantum, wiseman2010quantum, jacobs2010stochastic}, which describe the open-system dynamics of quantum optical systems interacting with an environment. QSDEs have been extensively studied in the context of quantum feedback and coherent control~\cite{wiseman1994quantum, belavkin1992quantum, James2008H, Nurdin2009Coherent, dong2010quantum,dong2023learning,dong2019learning}, and provide a natural formalism for modeling continuous-time evolution under both Hamiltonian and dissipative processes. Related Fr\'echet-derivative techniques under physical
Realizability (PR) constraints in coherent quantum control have been developed in
\cite{vladimirov2013quasi}, and Gaussian invariant-state descriptions based on
Wick--Isserlis relations can be found in \cite{vladimirov2012gaussian}.

Our framework models system dynamics in terms of annihilation and creation operators, making it especially suitable for bosonic platforms. By parameterizing the system's interaction terms, we minimize the steady-state energy as a variational cost function. Crucially, since accurately preparing a known initial state is often technically demanding \cite{preskill2018quantum}, we design the system to converge to a unique steady state regardless of initial conditions or perturbations. This eliminates the need for state reinitialization or repeated measurement. Such inherent stability enhances the scalability, robustness, and experimental feasibility of the proposed approach.

To further enhance robustness, we incorporate $H^\infty$ control techniques to mitigate persistent or structured noise. 
$H^\infty$ control is a strategy designed to minimize the worst-case impact of disturbances on system performance, ensuring guaranteed stability margins even under significant uncertainty~\cite{zhou1996robust,dong2023learning}. 
This robustness effectively suppresses environmental noise in quantum systems~\cite{James2008H, Nurdin2009Coherent, maalouf2009coherent, xiang2017coherent}.
Notably, Mabuchi~\cite{Mabuchi2008Coherent} has implemented such a control scheme in an all-optical experiment, validating its practical feasibility in photonic systems. 
This connection emphasizes the applied value of our approach and its suitability for implementation in existing quantum optical infrastructures.

We demonstrate the proposed method on the hydrogen molecule (H\textsubscript{2}), modeled by a second-quantized fermionic Hamiltonian mapped into the bosonic domain \cite{tilly2022variational}. The optical quantum system is parametrized and optimized using our variational framework, enabling steady-state energy minimization through direct adjustment of physically realizable system parameters. We compare the performance of our approach with the quantum approximate optimization algorithm (QAOA). In addition, we incorporate a robust $H^\infty$ controller \cite{James2008H} to illustrate the framework's resilience against external noise and perturbations, highlighting its potential for stable implementation in open quantum systems.

The remainder of the paper is organized as follows. Section~\ref{sec:qsde} outlines the construction of the QSDE model that governs the open-system dynamics of the quantum optical system. Section~\ref{sec:design} presents the variational framework used to optimize physically realizable parameters for steady-state energy minimization. In Section~\ref{sec:robust}, we extend the design to incorporate an $H^\infty$ controller, providing robustness against external disturbances. Section~\ref{sec:example} reports numerical results for ground-state energy computation of the hydrogen molecule, comparing our method with classical and quantum variational approaches, and demonstrating the impact of robust control. This work thus bridges the theory of quantum feedback control with practical ground-state computation, offering an alternative to conventional qubit-based methods.

\section{Quantum Stochastic Dynamics and Energy Objective}
\label{sec:qsde}

An open quantum system can be modeled through its Hamiltonian and coupling operators, following the formalism established in the quantum input-output theory of Gardiner and Collett~\cite{gardiner2004quantum}, and later systematized by Gough and James~\cite{Gough2009Series}. Consider a system described by a set of annihilation operators $\mathbf{a} = (\mathbf{a}_1, \ldots, \mathbf{a}_n)^\top$ acting on a Fock space. The evolution of the system and its environment is characterized by a triple $(\mathbf{S}, \mathbf{L}, \mathbf{H})$:
\begin{itemize}
    \item $\mathbf{H} = \mathbf{a}^\dagger \mathsf{\Omega} \mathbf{a}$ is a quadratic system Hamiltonian, where $\mathsf{\Omega} = \mathsf{\Omega}^\dagger \in \mathbb{C}^{n \times n}$, and $^\dagger$ denotes the Hermitian conjugate.
    \item $\mathbf{L} = \mathsf{C} \mathbf{a}$ is the coupling operator, with $\mathsf{C} \in \mathbb{C}^{m \times n}$,
    \item $\mathbf{S} = I$ is the scattering matrix, assumed to be identity.
\end{itemize}

The system evolution in the Heisenberg picture is governed by the Hudson-Parthasarathy quantum stochastic differential equation (QSDE)~\cite{hudson1984quantum}:
\begin{align}
    dU(t) = \left\{ -\left(i \mathbf{H} + \tfrac{1}{2} \mathbf{L}^\dagger \mathbf{L} \right)dt + \mathbf{L}^\dagger db(t) - \mathbf{L} db^\dagger(t) \right\} U(t),
\end{align}
where \( b(t) \) is the vector of bosonic input fields.

The Heisenberg evolution of the annihilation operator vector is:
\begin{align}
    d\mathbf{a}(t) = -i \mathsf{\Omega} \mathbf{a}(t) dt - \tfrac{1}{2} \mathsf{C}^\dagger \mathsf{C} \mathbf{a}(t) dt - \mathsf{C}^\dagger db(t).
\end{align}

Introducing the doubled-up notation:
\begin{align}
    \nu(t) = \begin{pmatrix} \mathbf{a}(t) \\ \mathbf{a}^\#(t) \end{pmatrix}, \quad dW(t) = \begin{pmatrix} db(t) \\ db^\#(t) \end{pmatrix},
\end{align}
where \( \mathbf{a}^\# \) denotes \( \mathbf{a}^\# = [\mathbf{a}_{ij}^*] \), and \( ^* \) denotes the adjoint operator or the complex conjugate, depending on the context. The system dynamics can be expressed as:
\begin{align}
    d\nu(t) = A \nu(t) dt + B dW(t)
\end{align}
with system matrices:
\begin{align}
    A &= \Delta(-i \mathsf{\Omega} - \tfrac{1}{2} \mathsf{C}^\dagger \mathsf{C}, 0), \\
    B &= -\Delta(\mathsf{C}^\dagger, 0),
\end{align}
where $\Delta(X, Y) = \begin{pmatrix} X & Y \\ Y^\# & X^\# \end{pmatrix}$ denotes the doubled-up operator.

The quantum noise satisfies:
\begin{align}
    \mathbb{E}[dW dW^\dagger] = F_w dt, \quad F_w = \begin{pmatrix} I & 0 \\ 0 & 0 \end{pmatrix}.
\end{align}

We consider the ground-state preparation problem for a cost Hamiltonian commonly arising in quantum chemistry~\cite{tilly2022variational}, given in second-quantized form:
\begin{align}
    H_{\mathrm{cost}} = \sum_i h_i \mathbf{a}_i^\dagger \mathbf{a}_i + \sum_{i,j} g_{ij} \mathbf{a}_i^\dagger \mathbf{a}_j^\dagger \mathbf{a}_j \mathbf{a}_i.
\end{align}
This Hamiltonian contains both quadratic and quartic bosonic interaction terms. We define the total steady-state energy cost as the sum of two terms:
\begin{align}
    \mathcal{J} := \mathcal{J}_1 + \mathcal{J}_2,
\end{align}
where the quadratic cost is given by:
\begin{align}
    \mathcal{J}_1 := \operatorname{Tr}(Q_1 \mathbf{S}_1),
\end{align}
and the quartic cost by:
\begin{align}
    \mathcal{J}_2 := \operatorname{Tr}(Q_2 \mathbf{S}_2).
\end{align}
Here, \( Q_1 \in \mathbb{C}^{2n \times 2n} \) and \( Q_2 \in \mathbb{C}^{(2n)^2 \times (2n)^2} \) are Hermitian cost matrices derived from the coefficients \( h_i \) and \( g_{ij} \), respectively.

The cost functionals \( \mathcal{J}_1 \) and \( \mathcal{J}_2 \) are determined by the steady-state statistics of the open quantum system. Assuming stability, the system reaches a unique steady state characterized by the first-order covariance:
\begin{align}
    \mathbf{S}_1 := \lim_{t \to \infty} \mathbb{E}[\nu(t) \nu^\dagger(t)],
\end{align}
and the second-order covariance:
\begin{align}
    \mathbf{S}_2 := \lim_{t \to \infty} \mathbb{E}[(\nu(t) \nu^\dagger(t)) \otimes (\nu(t) \nu^\dagger(t))].
\end{align}

\section{Variational Design under Physical Realizability Constraints}
\label{sec:design}

We present a variational framework for synthesizing linear quantum systems whose steady states approximate the ground states of second-quantized Hamiltonians. The approach proceeds by parameterizing the system matrices with a real-valued vector of tunable parameters and minimizing an energy objective derived from first- and second-order steady-state covariance matrices.

To ensure that the synthesized quantum system is physically realizable, we adopt a parameterization that enforces the physical realizability condition~\cite{James2008H}. Specifically, let the system matrices \( Y(\boldsymbol{\theta}) \in \mathbb{C}^{2n \times 2n} \) and \( B(\boldsymbol{\theta}) \in \mathbb{C}^{2n \times 2m} \) be expressed as linear combinations of fixed basis matrices:
\begin{align}
    Y(\boldsymbol{\theta}) := \sum_{i=1}^d \theta_i Y_i, \qquad
    B(\boldsymbol{\theta}) := \sum_{i=1}^d\sqrt{\theta_i} B_i.
\end{align}
Each pair \( (Y_i, B_i) \) is selected such that the corresponding system satisfies the structural constraints of quantum stochastic models, including the PR condition. The PR constraint is enforced by computing the drift matrix \( A(\boldsymbol{\theta}) \) using the following construction:
\begin{align}
    A(\boldsymbol{\theta}) := \frac{1}{2} \left( Y(\boldsymbol{\theta}) - Y(\boldsymbol{\theta})^\flat \right) - \frac{1}{2} B(\boldsymbol{\theta}) B(\boldsymbol{\theta})^\flat,
\end{align}
where the flat operation is defined as \( X^\flat := \mathbb{J} X^\dagger \mathbb{J} \), and the canonical structure matrix is \( \mathbb{J} = \begin{pmatrix} I & 0 \\ 0 & -I \end{pmatrix} \). This construction guarantees that 
\begin{align}
    A(\boldsymbol{\theta}) + A(\boldsymbol{\theta})^\flat + B(\boldsymbol{\theta}) B(\boldsymbol{\theta})^\flat = 0 , 
\end{align}
ensuring PR throughout the optimization process.

The system dynamics are governed by QSDEs, where the operators and noise matrices are linear combinations of physically realizable basis components:
\begin{align}
\label{eqn:QSDEs_theta}
    d\nu(t) = A(\boldsymbol{\theta}) \nu(t)\,dt + B(\boldsymbol{\theta}) dW(t).
\end{align}
The covariance matrices of the system in \eqref{eqn:QSDEs_theta} can be computed by solving algebraic Lyapunov equations (ALEs). For the first-order covariance matrix, the matrix \( \mathbf{S}_1 \) satisfies:

\begin{align}
\label{eqn: ALE}
    A(\boldsymbol{\theta})\, \mathbf{S}_1 + \mathbf{S}_1\, A^\dagger(\boldsymbol{\theta}) + B(\boldsymbol{\theta}) F_w B^\dagger(\boldsymbol{\theta}) = 0.
\end{align}
The second-order moment matrix \( \mathbf{S}_2 \) satisfies:
\begin{align}
\label{eq:second_order_lyapunov}
\begin{aligned}
     &(A \otimes I + I \otimes A)\, \mathbf{S}_2 + \mathbf{S}_2\, (A^\dagger \otimes I + I \otimes A^\dagger) \\
     &\quad + (B F_w B^\dagger) \otimes \mathbf{S}_1 + \mathbf{S}_1 \otimes (B F_w B^\dagger) + \mathsf{M} = 0.
\end{aligned}
\end{align}
The detailed derivation of these ALEs are provided in Appendix~\ref{appendix:ALE}, and the matrix \( \mathsf{M} \) is defined explicitly in \eqref{eqn:M_def}.

The goal of the optimization procedure is to minimize the cost function \( \mathcal{J}(\boldsymbol{\theta}) \) by adjusting the system parameters \( \boldsymbol{\theta} \), while ensuring that the resulting quantum system remains stable and physically realizable throughout the process.

The complete workflow is summarized in Algorithm~\ref{alg:variational}. At each iteration, the system matrices \( A(\boldsymbol{\theta}) \) and \( B(\boldsymbol{\theta}) \) are constructed from a parameterized basis. The steady-state covariance matrices are then computed via the ALEs, and the cost \( \mathcal{J}(\boldsymbol{\theta}) \) is evaluated accordingly. 

An important practical consideration is that, although solving ALEs incurs exponential computational complexity in large-scale systems, this difficulty is avoided in physical quantum platforms. In such systems, the state naturally evolves toward a steady state governed by the same underlying dynamics. Thus, the required covariance matrices can be directly accessed through measurement, enabling the proposed variational method to remain experimentally feasible even when classical simulation becomes intractable.

\begin{algorithm}[H]
\caption{Variational Optimization of $\mathcal{J}(\boldsymbol{\theta})$}
\label{alg:variational}
\begin{algorithmic}[1]
\State Initialize parameter $\boldsymbol{\theta}_0$
\For{each iteration $t = 1, 2, \dots$}
    \State Construct system matrices $Y(\boldsymbol{\theta}_t)$ and $B(\boldsymbol{\theta}_t)$
    \State Compute $A(\boldsymbol{\theta}_t) = \frac{1}{2}(Y(\boldsymbol{\theta}_t) - Y(\boldsymbol{\theta}_t)^\flat) - \frac{1}{2} B(\boldsymbol{\theta}_t) B(\boldsymbol{\theta}_t)^\flat$
    \State Evaluate $\mathcal{J}(\boldsymbol{\theta}_t)$ by measuring the steady-state moments \textit{(computed via ALEs \eqref{eqn: ALE}\eqref{eq:second_order_lyapunov} in simulation)}
    \State Update $\boldsymbol{\theta}_{t+1}$ using the chosen optimization method
\EndFor
\end{algorithmic}
\end{algorithm}

In this work, we adopt Simultaneous Perturbation Stochastic Approximation (SPSA)~\cite{spall1992multivariate} for the variational optimization due to its efficiency in high-dimensional systems. Specifically, at each iteration, SPSA updates the parameters $\boldsymbol{\theta}_t$ by generating a random perturbation vector $\Delta \in \{\pm 1\}^{d}$, where $d$ is the dimension of $\boldsymbol{\theta}$. The gradient is estimated using two cost evaluations:

\begin{align}
    g_t = \frac{\mathcal{J}(\boldsymbol{\theta}_t + c_t \Delta) - \mathcal{J}(\boldsymbol{\theta}_t - c_t \Delta)}{2 c_t \Delta}
\end{align}

The parameter is then updated according to

\begin{align}
    \boldsymbol{\theta}_{t+1} = \boldsymbol{\theta}_t - a_t g_t
\end{align}

Here, $a_t$ and $c_t$ are the learning rate and perturbation scale at iteration $t$, respectively, which are typically chosen to decay with $t$ to ensure convergence. This method is particularly suitable for high-dimensional systems, where gradient descent methods become computationally infeasible due to the exponential cost of gradient evaluation. In experimental scenarios where full quantum state access is unavailable, SPSA remains applicable as it requires only a few noisy cost evaluations per iteration, making it feasible for measurement-driven quantum optimization.

% In contrast, to establish a theoretical performance benchmark, we also implement a direct gradient descent procedure to optimize the system matrices \( A \) and \( B \) themselves. Explicit expressions for \( \partial \mathcal{J} / \partial A \) and \( \partial \mathcal{J} / \partial B \) are provided in Appendix~\ref{appendix:gradient}. Since the steady-state covariances \( \mathbf{S}_1 \) and \( \mathbf{S}_2 \) are deterministically computable from the system matrices, the optimization avoids measurement noise. However, this approach suffers from exponential computational complexity and is only tractable for small systems. It nonetheless serves as a useful reference for evaluating the performance of the variational parameterization approach.

In summary, the proposed variational framework enables the scalable and physically realizable synthesis of quantum systems that approximate ground states of general bosonic Hamiltonians using steady-state dynamics. A key advantage of this approach is its intrinsic robustness: the system converges to a unique steady state regardless of the initial condition and is capable of recovering from sudden perturbations without the need for reinitialization or measurement-based resets. These features make the framework particularly well suited for implementation on open quantum platforms subject to environmental noise. In the next section, we extend this formulation by incorporating an \( H^\infty \) controller to further enhance robustness against structured disturbances and performance degradation.

\section{Robust Performance via Coherent \( H^\infty \) Control}
\label{sec:robust}
While the proposed variational framework ensures convergence under ideal conditions, realistic quantum systems are inevitably subject to noise, model uncertainties, and external disturbances. To enhance robustness in such settings, we incorporate coherent \( H^\infty \) control~\cite{James2008H}, which generalizes classical \( H^\infty \) strategies to quantum systems and aims to minimize the worst-case gain from disturbance inputs to performance outputs~\cite{zhou1996robust}.

The synthesis of a coherent \( H^\infty \) controller involves solving a pair of algebraic Riccati equations~\cite{green2012linear,zhou1996robust}. However, due to PR constraints unique to quantum systems, a quantum version of the Riccati-based synthesis procedure has been proposed in~\cite{petersen1991first, James2008H} to ensure that the resulting controller can be physically implemented.

We consider an augmented quantum system driven by three types of inputs: vacuum noise \( dW(t) \), external disturbance \( dw(t) \), and coherent control input \( du(t) \). The system dynamics are given by
\begin{align}
    d\nu(t) &= A(\boldsymbol{\theta}) \nu(t)\, dt + B(\boldsymbol{\theta})\, dW(t) + B_1\, dw(t) + B_2\, du(t), \nonumber \\
    dz(t) &= C_1 \nu(t)\, dt + D_{12}\, du(t), \nonumber \\
    dy(t) &= C_2 \nu(t)\, dt + D_{20}\, dW(t) + D_{21}\, dw(t),
\end{align}
where \( dz(t) \in \mathbb{C}^{p_z} \) and \( dy(t) \in \mathbb{C}^{p_y} \) correspond to the performance output and the controller's measurement input, respectively, with \( p_z \) denoting the number of performance channels to be regulated and \( p_y \) the number of observed output channels. All system matrices \( A, B, B_1, B_2, C_1, C_2, D_{12}, D_{20}, D_{21} \) are appropriately dimensioned and satisfy the PR conditions~\cite{James2008H}.

The control objective is to synthesize a feedback controller that maps the measurement signal \( dy(t) \) to the control input \( du(t) \), ensuring that the resulting closed-loop system is internally stable and satisfies the robust performance condition
\begin{align}
    \| T_{dw \to dz} \|_\infty  < g,
\end{align}
for a prescribed attenuation level \( g > 0 \). Here, \( T_{dw \to dz} \) denotes the closed-loop transfer function from the disturbance input \( dw \) to the performance output \( dz \), and \( \| \cdot \|_\infty \) refers to the \( H^\infty \) norm \cite{zhou1996robust}.

Following~\cite{James2008H}, the controller is synthesized by solving a pair of coupled algebraic Riccati equations. When stabilizing solutions \( X \) and \( Y \) exist, the controller dynamics are given by
\begin{align}
\begin{aligned}
     d\xi(t) &= A_K \xi(t)\, dt + B_K\, dy(t) + B_{k1}\, dW_k(t), \\
     du(t) &= C_K \xi(t)\, dt + B_{k0}\, dW_k(t),
\end{aligned}
\end{align}
where \( \xi(t) \in \mathbb{C}^{2n} \) is the controller's internal state, and \( dW_k(t) \) is an independent vacuum noise process. The matrices \( B_{k0} \) and \( B_{k1} \) can be freely chosen. If the controller is required to be fully quantum (i.e., physically realizable as a quantum system), then these matrices must satisfy PR constraints. However, if a quantum-classical hybrid architecture is allowed—such as via measurement-based feedback—then the controller needs not adhere to PR conditions~\cite{James2008H}.

The controller matrices are constructed as
\begin{align}
    A_K &= A + B_2 C_K - B_K C_2 + (B_1 - B_K D_{21}) B_1^\top X, \nonumber \\
    B_K &= (I - YX)^{-1} (Y C_2^\top + B_1 D_{21}^\top) E_2^{-1}, \nonumber \\
    C_K &= - E_1^{-1} (g^2 B_2^\top X + D_{12}^\top C_1),
\end{align}
with \( E_1 = D_{12}^\top D_{12} \), \( E_2 = D_{21} D_{21}^\top \), and \( g > 0 \) denoting the prescribed disturbance attenuation level, the Riccati equations used to compute \( X \) and \( Y \) are given in \cite[Eqs. (27) and (28)]{James2008H}. This controller guarantees that the closed-loop system is internally stable and satisfies the robust performance condition \( \| T_{dw \to dz} \|_\infty < g \).

The overall closed-loop system, comprising both the plant and controller, is described by
\begin{align}
    d\eta(t) &= \tilde{A} \eta(t)\, dt + \tilde{B}\, dw(t) + \tilde{G}\, d\tilde{W}(t), \nonumber \\
    dz(t) &= \tilde{C} \eta(t)\, dt + \tilde{H}\, d\tilde{W}(t),
\end{align}
where \( \eta(t) = \begin{bmatrix} \nu(t)^\top & \xi(t)^\top \end{bmatrix}^\top \) is the augmented system state, and \( d\tilde{W}(t) = \begin{bmatrix} dW(t)^\top & dW_k(t)^\top \end{bmatrix}^\top \) aggregates the vacuum noise processes. The closed-loop matrices are defined as
\begin{align}
\label{eqn:closed-form}
    \tilde{A} &= \begin{bmatrix}
        A & B_2 C_K \\
        B_K C_2 & A_K
    \end{bmatrix}, \quad
    \tilde{B} = \begin{bmatrix}
        B_1 \\
        B_K D_{21}
    \end{bmatrix}, \nonumber \\
    \tilde{G} &= \begin{bmatrix}
        B & 0 \\
        B_K D_{20} & B_{k1}
    \end{bmatrix}, \quad
    \tilde{C} = \begin{bmatrix}
        C_1 & D_{12} C_K
    \end{bmatrix}, \nonumber \\
    \tilde{H} &= \begin{bmatrix}
        0 & D_{12} B_{k0}
    \end{bmatrix}.
\end{align}

\begin{figure}[ht]
\centering
\begin{tikzpicture}[auto, node distance=2.2cm and 2cm, >=Latex, every node/.style={align=center}]
% \begin{tikzpicture}[scale=1, every node/.style={font=\small}]
  % Define key points
  \coordinate (k1) at (0, 3);
  \coordinate (k2) at (4, 3);
  \coordinate (k3) at (2, 1);
  \coordinate (a) at (2, 2.2);

  % Plant boundary
  \draw[dashed] (-1,0.5) rectangle (5,3.5);
  \node at (2,2.2) {plant};
  % Beam splitters
  \draw[very thick, rotate around={45:(k1)}] ($(k1)+(-0.6,0)$) -- ++(1.2,0) node[midway, above left=1pt] {};
  \draw[very thick, rotate around={-45:(k2)}] ($(k2)+(-0.6,0)$) -- ++(1.2,0) node[midway, above right=1pt] {};
  \draw[very thick] ($(k3)+(-0.7,0)$) -- ++(1.4,0) node[midway, below=2pt] {};

  % Mode a connections with spacing
  \draw[->] ($(k3)+(-0.3,0.3)$) -- ($(k1)+(0.15,-0.15)$);
  \draw[->] ($(k2)+(-0.15,-0.15)$) -- ($(k3)+(0.3,0.3)$);
  \draw[->] ($(k1)+(0.3,0)$) -- ($(k2)+(-0.3,0)$) node[midway, above] {};

  % Input v
  \draw[->] (-1.3,3) -- ($(k1)+(-0.2,0)$) node[above left] {\(dW\)};
  \draw[->] ($(k1)+(0,0.2)$) -- ++(0,1.5);

  % Input w
  \draw[<-] ($(k2)+(0,0.2)$) -- ++(0,1.5) node[midway, right] {\(dw\)};

  % Output y
  % \coordinate (yout) at ($(k2)+(2,0)$);
  % \draw[->] ($(k2)+(0.2,0)$) -- (yout) node[midway, above] {\(y\)};

  % Controller block
  \node[draw, rectangle, minimum width=3.2cm, minimum height=1.2cm, fill=green!10, anchor=north] (controller) at (4.5,-0.3)  {Coherent Controller};

  % Arrow from y to controller (bent)
  \draw[->] 
    ($(k2)+(0.2,0)$) -- ++(1.5,0) node[above right] {\(y\)}
           -- ++(0,-3.2);
           % -- ($(controller.north)$);

  % Output z
  \draw[->] ($(k3)+(-0.1,-0.2)$) -- ++(-1,-1) node[above left] {\(z\)};

  % Controller output to u
  \draw[<-] ($(k3)+(0.1,-0.2)$) -- ++(1,-1) node[above right] {\(u\)};

\end{tikzpicture}
\caption{Closed-loop quantum \( H^\infty \) control architecture with coherent feedback.}
\label{fig:hinf-structure}
\end{figure}
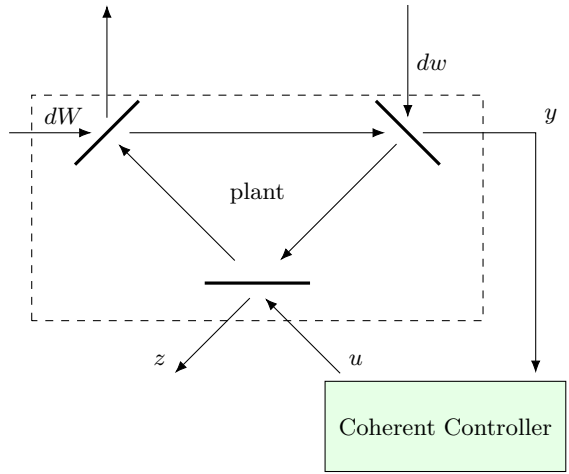

The overall feedback architecture is illustrated in Fig.~\ref{fig:hinf-structure}, where coherent feedback enables robustness without classical post-processing.

% \begin{figure}[ht]
% \centering
% \begin{tikzpicture}[auto, node distance=2.2cm and 2cm, >=Latex, every node/.style={align=center}]
% \node[draw, rectangle, minimum width=2.5cm, minimum height=1.2cm, fill=blue!10] (plant) {Quantum Plant};
% \node[draw, rectangle, minimum width=2.5cm, minimum height=1.2cm, fill=green!10, below=2cm of plant] (controller) {Coherent Controller};

% \node[left=1.5cm of plant] (input) {$[dW, dw]$};
% \node[right=2.8cm of plant] (zout) {$z$};

% \node[left=1.5cm of controller] (control input) {$dW_k$};

% \draw[->] (input) -- (plant.west);
% \draw[->] (control input) -- (controller.west);
% \draw[->] (plant.east) -- (zout);
% \draw[->] ([xshift=0.3cm]controller.north) -- node[right] {$u$} ([xshift=0.3cm]plant.south);
% \draw[->] ([xshift=-0.3cm]plant.south) -- node[left] {$y$} ([xshift=-0.3cm]controller.north);
% \end{tikzpicture}
% \caption{Closed-loop quantum \( H^\infty \) control architecture with coherent feedback.}
% \label{fig:hinf-structure}
% \end{figure}

This integration of coherent \( H^\infty \) control not only enhances resilience to disturbances but also naturally complements the variational steady-state design. In particular, once the quantum plant matrices \( A \) and \( B \) are obtained through the variational procedure (Algorithm~\ref{alg:variational}), they can be directly used to synthesize an \( H^\infty \) controller tailored to the system dynamics. This enables the immediate deployment of robust control to mitigate external perturbations without requiring structural redesign. The performance of the resulting \( H^\infty \) controller under disturbance is demonstrated in the numerical simulations presented in Section~\ref{sec:example}.

\section{Application to Ground-State Preparation of H\textsubscript{2}}
\label{sec:example}

To demonstrate the effectiveness of our framework, we apply it to the ground-state energy estimation of the hydrogen molecule (H\textsubscript{2}). In this section, we first describe the modeling of the H\textsubscript{2} Hamiltonian, and then compare three computational methods—QAOA, matrix-level gradient descent, and our variational approach—before evaluating robustness through \( H^\infty \) control.

\subsection{Modeling of the H\textsubscript{2} Hamiltonian}

The system is modeled using a second-quantized Hamiltonian that captures both single-particle and interaction terms~\cite{tilly2022variational}:
\begin{align}
    H_{\text{cost}} = c + \sum_i h_i\, \mathbf{a}_i^\dagger \mathbf{a}_i + \sum_{i,j} g_{ij}\, \mathbf{a}_i^\dagger \mathbf{a}_j^\dagger \mathbf{a}_j \mathbf{a}_i,
\end{align}
where \( c \in \mathbb{R} \) is a constant energy offset, \( h_i \) are the single-particle coefficients, and \( g_{ij} \) represent quartic interaction strengths. These coefficients are obtained from a fermionic second-quantized Hamiltonian using \texttt{OpenFermion}~\cite{mcclean2020openfermion} and subsequently mapped to the bosonic domain under the assumption of Gaussian-preserving dynamics.

\subsection{Method I: Quantum Approximate Optimization Algorithm (QAOA)}

As a baseline comparison, we implement the QAOA using the \texttt{Qiskit} framework~\cite{ibm_qiskit_2024}. To enable compatibility with QAOA's spin-based ansatz, the second-quantized fermionic Hamiltonian of H\textsubscript{2} is first transformed into a spin representation via the Jordan--Wigner transformation, as implemented in \texttt{OpenFermion}~\cite{mcclean2020openfermion}. This yields an effective Ising-type Hamiltonian expressed in terms of Pauli operators, which serves as the cost Hamiltonian for QAOA.

\begin{figure}[t]
    \centering
    \includegraphics[width=0.48\textwidth]{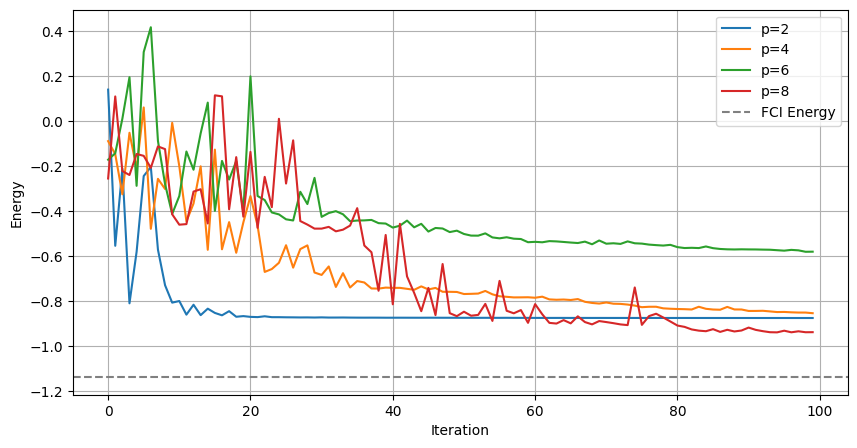}
    \caption{Convergence of QAOA with increasing circuit depth $p$ applied to the mapped hydrogen molecule Hamiltonian. FCI energy is shown as a dashed line.}
    \label{fig:qaoa}
\end{figure}

We then apply QAOA with increasing circuit depth \( p \), optimizing the variational parameters using COBYLA and adopting standard layer-wise mixer Hamiltonians. Fig.~\ref{fig:qaoa} shows the convergence behavior of the algorithm, where the full configuration interaction (FCI) energy is plotted as a reference. As expected, deeper circuits yield lower energy estimates, but at the cost of increased gate complexity and optimization overhead.

\subsection{Method II: Matrix-Level Gradient Optimization}

To validate our variational framework, we simulate a truncated model of the hydrogen molecule, considering only the dominant contributions from two spin-orbitals. The cost Hamiltonian is derived by projecting the full molecular Hamiltonian onto this two-mode subspace, resulting in \cite{mcclean2020openfermion}:
\begin{align}
\begin{aligned}
    H_{\text{cost}} &= 0.715104 - 1.253310 \sum_{i=0}^{1} a_i^\dagger a_i \\
    &+ 0.337378 \sum_{i,j=0}^{1} a_i^\dagger a_i^\dagger a_j a_j,
\end{aligned}
\end{align}
where \( a_i^\dagger \), \( a_i \) are bosonic creation and annihilation operators. This simplified model admits a theoretical minimum energy of approximately \( -1.1168 \) Hartree.

As a theoretical benchmark, we also compute the minimal cost by treating the system matrices \( A \) and \( B \) as unconstrained (but physically realizable) variables and directly minimizing the steady-state cost \( \mathcal{J} \) subject to Lyapunov equations. The gradient of \( \mathcal{J} \) is analytically derived in Appendix~\ref{appendix:gradient}. This approach yields a lowest achievable value of \( -1.1572 \), which serves as a lower bound for all linear quantum realizations of the truncated model.

\begin{figure}[t]
    \centering
    \includegraphics[width=0.48\textwidth]{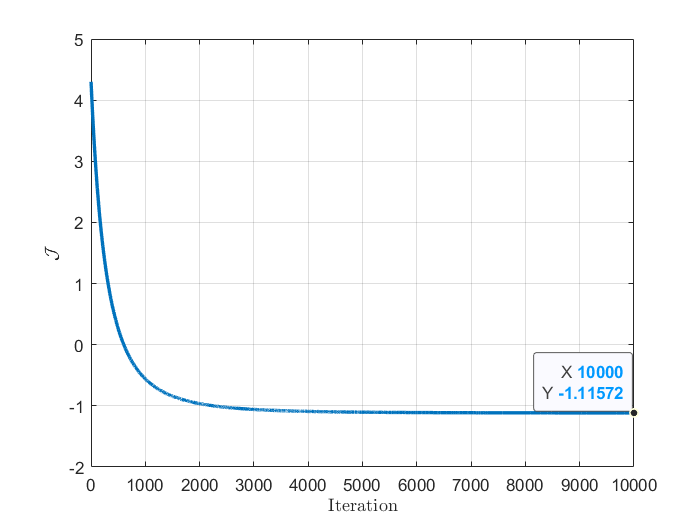}
    \caption{Cost function $\mathcal{J}(\theta)$ over iterations of gradient-based optimization applied directly to system matrices $A$ and $B$.}
    \label{fig:gd-cost}
\end{figure}

\begin{figure}[t]
    \centering
    \includegraphics[width=0.48\textwidth]{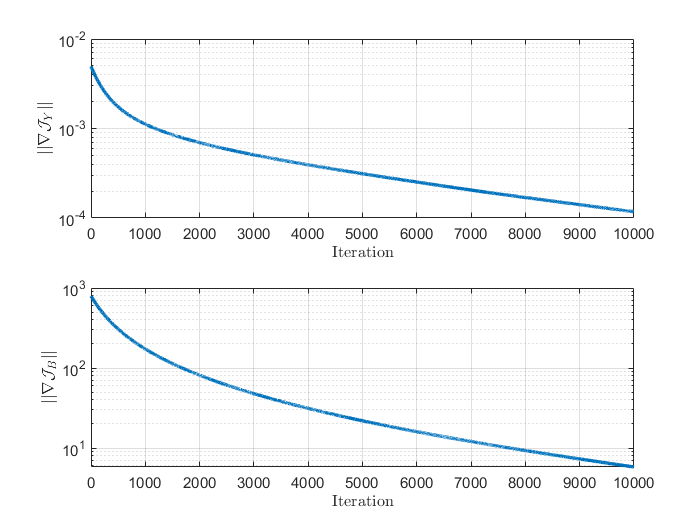}
    \caption{Norms of gradients $\|\nabla \mathcal{J}_Y\|$ and $\|\nabla \mathcal{J}_B\|$ during matrix-level QSDE optimization. Convergence of gradients confirms optimality of the resulting quantum system.}
    \label{fig:gd-grad}
\end{figure}

In comparison, the FCI value for the original hydrogen model is \( -1.1373 \) Hartree. Our variational method, relying only on second-order moments and physical realizability constraints, achieves an energy remarkably close to the FCI result, demonstrating the effectiveness of our approach. The convergence behavior is illustrated in Figs.~\ref{fig:gd-cost} and~\ref{fig:gd-grad}.

\subsection{Method III: Variational QSDE-Based Optimization}

We now apply our proposed variational strategy (Algorithm~\ref{alg:variational}), where the parameter vector \( \boldsymbol{\theta} \in \mathbb{R}^n \) is defined using a basis expansion. In this work, we select \( n \in \{2, 4, 6, 8\} \). At each iteration, the Lyapunov equation is solved to compute the steady-state covariance, and the corresponding cost function \( \mathcal{J}(\boldsymbol{\theta}) \) is evaluated. The optimization is performed using SPSA, with parameters set as follows:
\begin{itemize}
\item[-] Initial step size: \( a_0 = 1 \times 10^{-4} \)
\item[-] Initial perturbation size: \( c_0 = 1 \times 10^{-4} \)
\item[-] Learning rate decay: \( a_t = \frac{a_0}{(t + A)^{0.602}} \)
\item[-] Perturbation decay: \( c_t = \frac{c_0}{(t + 1)^{0.5}} \)
\end{itemize}

\begin{figure}[t]
    \centering
    \includegraphics[width=0.48\textwidth]{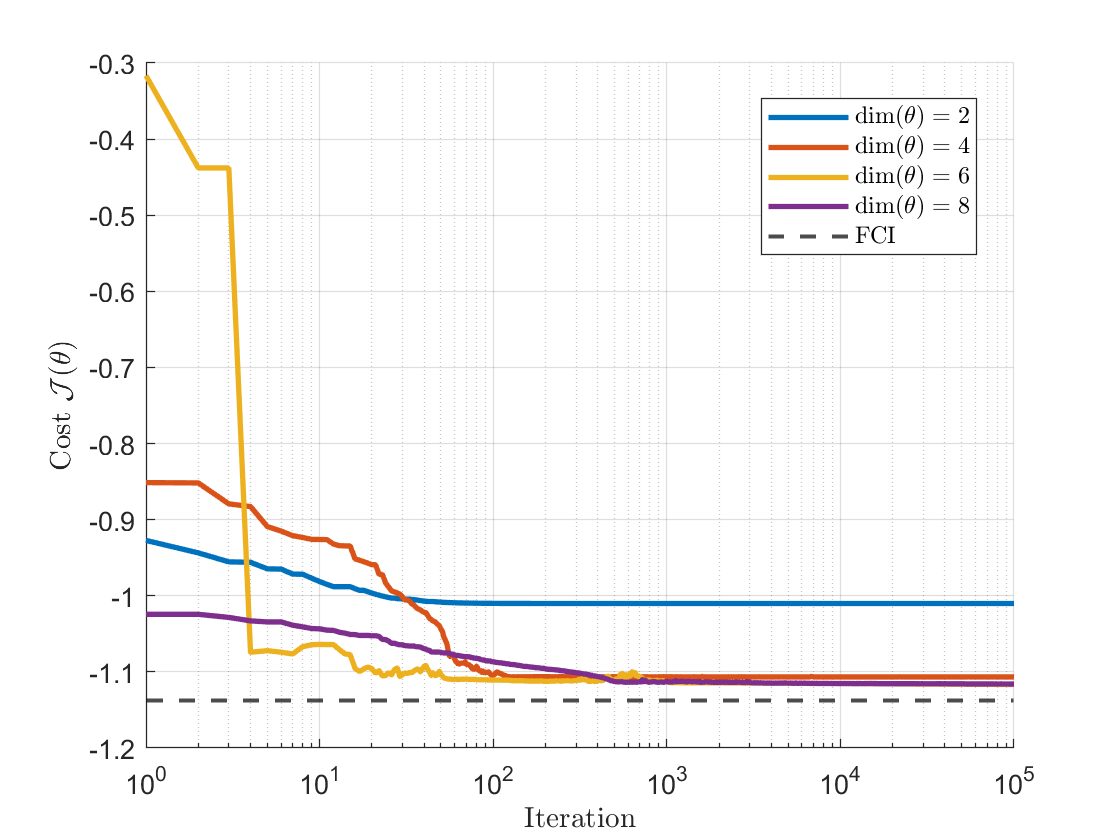}
    \caption{Variational optimization using the SPSA algorithm. The cost $\mathcal{J}(\theta)$ decreases steadily toward a ground energy.}
    \label{fig:SPSA}
\end{figure}

Although the full procedure is carried out classically in simulation, the underlying method is designed to be implementable on optical platforms, where steady-state covariances can be directly measured. Fig.~\ref{fig:SPSA} shows the convergence behavior of the cost function under the SPSA algorithm.

\subsection{Robust Performance under Disturbance}

An important advantage of our framework is its compatibility with robust control techniques. In particular, the integration of an \( H^\infty \) controller~\cite{James2008H} enables the system to maintain performance in the presence of external disturbances. To quantify robustness, we introduce a disturbance input \( dw(t) \) via the channel \( B_1 \). In simulation, we vary the disturbance scale by setting \( B_1 = -\alpha I \), where \( \alpha \in [1, 10] \), and define \( B_2 = -I \). The corresponding measurement matrix \( C_2, C_1 \) is chosen to satisfy physical realizability conditions, and we set \( D_{12} = D_{21} = I \). We then evaluate the resulting steady-state cost under both open-loop and closed-loop configurations to assess the impact of robust control.

The perturbed system is modeled as
\begin{align}
    d\nu(t) = A \nu(t)\, dt + B_0\, dW(t) + B_1\, dw(t),
\end{align}
where \( dW(t) \) denotes the vacuum input and \( dw(t) \) represents an exogenous disturbance. In the open-loop case, the system lacks active rejection of \( dw(t) \), resulting in increased steady-state cost as the disturbance norm grows. In contrast, the closed-loop configuration, implemented via the coherent \( H^\infty \) controller derived in Section~\ref{sec:robust}, effectively mitigates the disturbance. The controller dynamics follow the realization in ~\eqref{eqn:closed-form}.

\begin{figure}[t]
    \centering
    \includegraphics[width=0.48\textwidth]{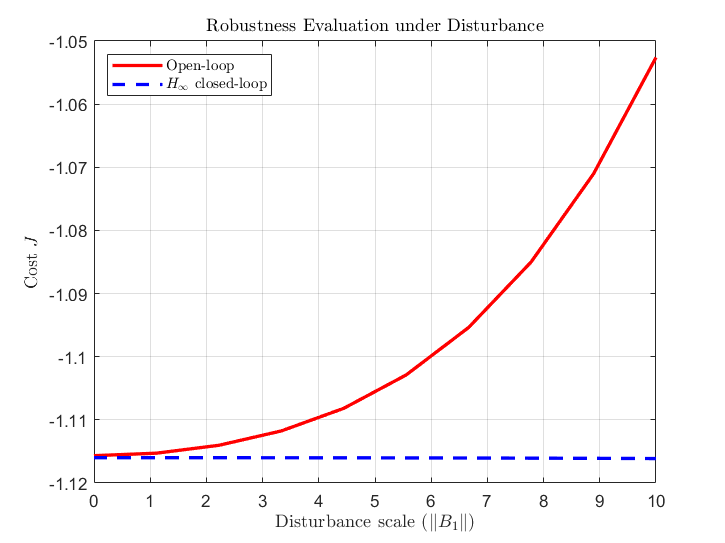}
    \caption{Robustness analysis under increasing disturbance intensity. The open-loop system exhibits rising cost, while the \( H^\infty \)-controlled system maintains bounded performance.}
    \label{fig:hinf-robustness}
\end{figure}

Fig.~\ref{fig:hinf-robustness} presents a quantitative comparison. The open-loop system exhibits a rapid rise in cost with increasing disturbance intensity, indicating poor robustness. In contrast, the \( H^\infty \)-controlled system maintains a nearly constant cost across the entire disturbance range, demonstrating effective noise attenuation. These results confirm the ability of our method to preserve performance under uncertainty, a key requirement for realistic quantum optical implementations.

\section{Conclusion}
\label{sec:conclusion}
We have introduced a physically motivated variational framework for computing ground states of quantum Hamiltonians by using the dissipative dynamics of open quantum systems governed by QSDEs. Our method operates directly in the bosonic domain, naturally aligning with the structure of quantum optical systems and eliminating the need for qubit-based encoding. By parameterizing the system's interactions and minimizing the steady-state energy, we enable ground-state preparation through stable, physically realizable evolution. To enhance robustness under external disturbances, we further incorporated coherent \( H^\infty \) control into the design. Numerical simulations on the hydrogen molecule demonstrate the effectiveness and scalability of the proposed approach, as well as its resilience to noise when compared to existing methods such as QAOA. Our results suggest a viable path toward implementing variational quantum algorithms in analog photonic systems, opening new opportunities for scalable and noise-tolerant quantum simulation.

\section*{Acknowledgment}
The authors thank Dr Igor G. Vladimirov for helpful discussions. We also acknowledge his
prior contributions to infinitesimal-perturbation and Fr\'echet-derivative methods for
coherent quantum control and filtering under physical realisability constraints, which
provide important background for the derivative calculations used in this work
\cite{vladimirov2013quasi}.

\appendix

\section{Algebraic Lyapunov Equations for Covariance Matrices}
\label{appendix:ALE}

To evaluate the cost functional \( \mathcal{J}(\boldsymbol{\theta}) \), we require both the first- and second-order steady-state moments of the quantum state. These moments can be obtained by solving corresponding ALEs under the assumption that the system is asymptotically stable.

We now derive the ALE that governs the first-order steady-state moment of the quantum system. Define the steady-state covariance matrix as
\begin{align}
    \mathbf{S}_1 := \lim_{t \to \infty} \mathbb{E}[\nu(t) \nu^\dagger(t)].
\end{align}

To determine the evolution of \( \mathbf{S}_1 \), we apply Itô’s rule to the matrix-valued process \( \nu(t)\nu^\dagger(t) \). Using the quantum stochastic differential equation \( d\nu = A \nu\,dt + B\,dw \), we obtain:
\begin{align}
\label{eq:ito-diff}
    d(\nu \nu^\dagger) = (d\nu)\, \nu^\dagger + \nu\, (d\nu)^\dagger + (d\nu)(d\nu)^\dagger.
\end{align}
Substituting for \( d\nu \), we expand:
\begin{align}
\begin{aligned}
    d(\nu \nu^\dagger) &= (A \nu\,dt + B\,dw)\, \nu^\dagger + \nu\, (A \nu\,dt + B\,dw)^\dagger \\
    &\quad + (A \nu\,dt + B\,dw)(A \nu\,dt + B\,dw)^\dagger.
\end{aligned}
\end{align}

Taking the expectation on both sides, and using the Itô table \( \mathbb{E}[dw] = 0 \), \( \mathbb{E}[dw\, dw^\dagger] = F_w\,dt \), the expectation of \eqref{eq:ito-diff} simplifies to:
\begin{align}
    \frac{d}{dt} \mathbb{E}[\nu \nu^\dagger] = A\, \mathbb{E}[\nu \nu^\dagger] + \mathbb{E}[\nu \nu^\dagger] A^\dagger + B F_w B^\dagger.
\end{align}
In the steady-state limit, \( \frac{d}{dt} \mathbf{S}_1 = 0 \), it yields the ALE for $S_1$:
\begin{align}
% \label{eqn: ALE}
    A(\boldsymbol{\theta})\, \mathbf{S}_1 + \mathbf{S}_1\, A^\dagger(\boldsymbol{\theta}) + B(\boldsymbol{\theta}) F_w B^\dagger(\boldsymbol{\theta}) = 0.
\end{align}

To capture the quartic contributions in the cost Hamiltonian, we derive the ALE for the second-order moment matrix
\begin{align}
    \mathbf{S}_2 := \lim_{t \to \infty} \mathbb{E}[(\nu \nu^\dagger) \otimes (\nu \nu^\dagger)].
\end{align}

To obtain the dynamics of \( \mathbf{S}_2 \), we apply Itô's rule to the matrix-valued process \( \nu \nu^\dagger \). From the QSDE \( d\nu = A \nu\,dt + B\,dw \), we compute the differential
\begin{align}
    d(\nu \nu^\dagger) = (d\nu) \nu^\dagger + \nu (d\nu)^\dagger + (d\nu)(d\nu)^\dagger,
\end{align}
and take its tensor product to obtain the evolution of \( \mathbf{S}_2 \):
\begin{align}
\label{eqn:dS2_tensor}
\begin{aligned}
    d\mathbf{S}_2 ={} &\mathbb{E}\left[ d(\nu \nu^\dagger) \otimes (\nu \nu^\dagger) \right] 
    + \mathbb{E}\left[ (\nu \nu^\dagger) \otimes d(\nu \nu^\dagger) \right] \\
    &+ \mathbb{E}\left[ d(\nu \nu^\dagger) \otimes d(\nu \nu^\dagger) \right].
\end{aligned}
\end{align}

We now evaluate each term in Eq.~\eqref{eqn:dS2_tensor}. First, expand
\begin{align}
\begin{aligned}
    d(\nu \nu^\dagger) &= (A \nu\,dt + B\,dw)\, \nu^\dagger + \nu\, (A \nu\,dt + B\,dw)^\dagger \\
    &\quad + (A \nu\,dt + B\,dw)(A \nu\,dt + B\,dw)^\dagger.
\end{aligned}
\end{align}

Using the quantum Itô rule \( \mathbb{E}[dw] = 0 \), \( \mathbb{E}[dw\, dw^\dagger] = F_w dt \), and taking expectations, we obtain:
\begin{align}
\begin{aligned}
    \mathbb{E}\left[d(\nu \nu^\dagger) \otimes (\nu \nu^\dagger)\right] 
    &= (A \otimes I)\, \mathbf{S}_2 + \mathbf{S}_2\, (A^\dagger \otimes I) \\
    &+ (B F_w B^\dagger) \otimes \mathbf{S}_1, \\
    \mathbb{E}\left[(\nu \nu^\dagger) \otimes d(\nu \nu^\dagger)\right] 
    &= (I \otimes A)\, \mathbf{S}_2 + \mathbf{S}_2\, (I \otimes A^\dagger) \\
    &+ \mathbf{S}_1 \otimes (B F_w B^\dagger).
\end{aligned}
\end{align}

The higher-order correction arising from quantum noise in the second-order moment equation is captured by the term
\begin{align}
\label{eqn:M_def}
    \mathsf{M} := \mathbb{E}[d(\nu \nu^\dagger) \otimes d(\nu \nu^\dagger)],
\end{align}
which can be decomposed as \( \mathsf{M} = \mathsf{M}_1 + \mathsf{M}_2 + \mathsf{M}_3 + \mathsf{M}_4 \), with each component expressible in terms of the first-order covariance matrix \( \mathbf{S}_1 \). We now derive this expression explicitly.

Using the quantum Itô rule and neglecting higher-order differentials \( (dt^2, dt \cdot dw, dw^3) \), the only non-vanishing contributions to \( \mathsf{M} \) are those containing exactly two increments \( dw \). Hence, we obtain:
\begin{align}
\begin{aligned}
    d(\nu \nu^\dagger) \otimes d(\nu \nu^\dagger)
    &= (B\,dw\,\nu^\dagger) \otimes (B\,dw\,\nu^\dagger)\\
    &+ (B\,dw\,\nu^\dagger) \otimes (\nu (dw)^\dagger B^\dagger) \\
    &\quad + (\nu (dw)^\dagger B^\dagger) \otimes (B\,dw\,\nu^\dagger)\\
    &+ (\nu (dw)^\dagger B^\dagger) \otimes (\nu (dw)^\dagger B^\dagger).
\end{aligned}
\end{align}

Taking expectations, and applying the quantum Markov property (which implies \( dw \) is independent of \( \nu \)), we use the identities:
\[
\mathbb{E}[dw\, dw^\dagger] = F_w dt, \qquad \mathbb{E}[dw\, \nu^\dagger] = 0,
\]
to obtain:
\begin{align}
\begin{aligned}
    \mathsf{M} ={} & (B \otimes B)\, \mathbb{E}[dw\, \nu^\dagger \otimes dw\, \nu^\dagger] \\
    &+ (B \otimes I)\, \mathbb{E}[dw\, \nu^\dagger \otimes \nu\, dw^\dagger] (I \otimes B^\dagger) \\
    &+ (I \otimes B)\, \mathbb{E}[\nu\, dw^\dagger \otimes dw\, \nu^\dagger] (B^\dagger \otimes I)\\
    &+ \mathbb{E}[\nu\, dw^\dagger \otimes \nu\, dw^\dagger] (B^\dagger \otimes B^\dagger).
\end{aligned}
\end{align}

To express each term in terms of \( \mathbf{S}_1 \), we define
\[
\mathbf{S}_1 = \left( \mathbf{s}_1\, \cdots\, \mathbf{s}_{2n} \right),
\quad \text{with } \mathbf{s}_i \in \mathbb{C}^{2n},
\]
and introduce a swapping matrix
\[
\mathbb{F} := \begin{pmatrix} 0 & I_n \\ I_n & 0 \end{pmatrix},
\]
which swaps the upper and lower components of each \( \mathbf{s}_i \).

We now define the correction matrices \( \mathsf{M}_i \in \mathbb{C}^{(2n)^2 \times (2n)^2} \) as follows:

\paragraph*{Term \(\mathsf{M}_1\):}
The matrix \( \mathsf{M}_1 \in \mathbb{C}^{(2n)^2 \times (2n)^2} \) is row-sparse. For each row index \( i \in \{1, \dots, (2n)^2\} \), define an index \( i' \in \{1, \dots, n\} \) such that
\begin{align}
    i = (2n)(i' - 1) + n + i'.
\end{align}
Then, for those specific \( i \), the corresponding row of \( \mathsf{M}_1 \) is
\begin{align}
    \mathsf{M}_1(i,:) = 
    \left( \mathbf{s}_{n+1}^\dagger \; \cdots \; \mathbf{s}_{2n}^\dagger \; \mathbf{s}_1^\dagger \; \cdots \; \mathbf{s}_n^\dagger \right).
\end{align}
All other rows of \( \mathsf{M}_1 \) are zero.

\paragraph*{Term \(\mathsf{M}_2\):}
The matrix \( \mathsf{M}_2 \in \mathbb{C}^{(2n)^2 \times (2n)^2} \) is defined as
\begin{align}
    \mathsf{M}_2 =
    \begin{pmatrix}
        \widetilde{M}_2 \\
        \mathbf{0}_{2n^2 \times (2n)^2}
    \end{pmatrix},
\end{align}
where \( \widetilde{M}_2 \in \mathbb{C}^{2n^2 \times (2n)^2} \) is a horizontally stacked block matrix, constructed in two segments:

\begin{itemize}
    \item For \( j = 0, \dots, n-1 \), the first \( n \) nonzero blocks are:
    \[
    \left[
        I_n \otimes (\mathbb{F} \mathbf{s}^*_{n+1+j}) \quad
        I_n \otimes \mathbf{0}_{2n \times 1}
    \right]
    \]
    \item For \( j = 0, \dots, n-1 \), the next \( n \) nonzero blocks are:
    \[
    \left[
        I_n \otimes (\mathbb{F} \mathbf{s}^*_{1+j}) \quad
        I_n \otimes \mathbf{0}_{2n \times 1}
    \right]
    \]
\end{itemize}

\paragraph*{Term \(\mathsf{M}_3\):}
The matrix \( \mathsf{M}_3 \) is block-wise sparse and given by
\begin{align}
    \mathsf{M}_3 =
    \begin{pmatrix}
        \mathbf{0}_{(2n)^2 \times 2n^2} & \widetilde{M}_3
    \end{pmatrix},
\end{align}
where \( \widetilde{M}_3 \in \mathbb{C}^{(2n)^2 \times 2n^2} \) is vertically stacked:
\begin{align}
    \widetilde{M}_3 =
    \begin{pmatrix}
        I_n \otimes \mathbf{0}_{1 \times 2n} \\
        I_n \otimes \mathbf{s}_1^\dagger \\
        I_n \otimes \mathbf{0}_{1 \times 2n} \\
        I_n \otimes \mathbf{s}_2^\dagger \\
        \vdots \\
        I_n \otimes \mathbf{0}_{1 \times 2n} \\
        I_n \otimes \mathbf{s}_{2n}^\dagger
    \end{pmatrix}.
\end{align}

\paragraph*{Term \(\mathsf{M}_4\):}
The matrix \( \mathsf{M}_4 \in \mathbb{C}^{(2n)^2 \times (2n)^2} \) is column-sparse. For each \( i' \in \{1, \dots, n\} \), define a column index
\begin{align}
    j = 2n^2 + (i' - 1)(2n) + i'.
\end{align}
Then, the corresponding column of \( \mathsf{M}_4 \) is given by
\begin{align}
    \mathsf{M}_4(:, j) =
    \begin{pmatrix}
        \mathbb{F} \mathbf{s}_1^* \\
        \mathbb{F} \mathbf{s}_2^* \\
        \vdots \\
        \mathbb{F} \mathbf{s}_{2n}^*
    \end{pmatrix}.
\end{align}
All other columns of \( \mathsf{M}_4 \) are zero.

\medskip
Together, the correction term 
\(
    \mathsf{M} = \mathsf{M}_1 + \mathsf{M}_2 + \mathsf{M}_3 + \mathsf{M}_4
\)
captures the quantum noise contribution in the second-order Lyapunov equation:
\begin{align}
% \label{eq:second_order_lyapunov}
\begin{aligned}
     &(A \otimes I + I \otimes A)\, \mathbf{S}_2 + \mathbf{S}_2\, (A^\dagger \otimes I + I \otimes A^\dagger) \\
     &\quad + (B F_w B^\dagger) \otimes \mathbf{S}_1 + \mathbf{S}_1 \otimes (B F_w B^\dagger) + \mathsf{M} = 0.
\end{aligned}
\end{align}
This equation provides the foundation for accurately computing the quartic cost contribution in the variational framework.

\begin{remark}(Gaussian invariant state and Wick--Isserlis relations)
For stable linear quantum stochastic systems driven by vacuum fields, a Gaussian
invariant state exists and higher-order moments can be expressed in terms of the first
two moments via Wick--Isserlis relations; see \cite{vladimirov2012gaussian}. The derivations in this appendix are included for completeness in our notation for the
Gaussian setting. Related moment-dynamics results for quasilinear systems
can be found in \cite{vladimirov2012characterization}.
\end{remark}

\section{Fr\'echet Derivatives with Respect to System Matrices}
\label{appendix:gradient}

We present the Fr\'echet-derivative expressions for the total cost
\( \mathcal{J} = \mathcal{J}_1 + \mathcal{J}_2 \)
with respect to the PR system matrices, using the infinitesimal-perturbation
framework developed in \cite{vladimirov2013quasi}. To enforce the PR constraint,
we parametrize the system matrix \( A \) as
\begin{align}
    A = -\tfrac{1}{2} B B^\flat + \frac{Y - Y^\flat}{2},
\end{align}
where \( Y \in \mathbb{C}^{2n \times 2n} \) and \( B \in \mathbb{C}^{2n \times 2m} \).
This parametrization guarantees that \( A + A^\flat + B B^\flat = 0 \) is always satisfied.

We define the costate matrix \( \Pi \) as the solution to the adjoint equation:
\begin{align}
    A^\dagger \Pi + \Pi A + Q = 0.
\end{align}
We further define the matrix:
\begin{align}
    \Omega := \Pi \mathbf{S}_1.
\end{align}

With these definitions, the Fr\'echet derivatives of the first-order cost term \( \mathcal{J}_1 = \mathrm{Tr}(Q \mathbf{S}_1) \) are given by:
\begin{align}
    \frac{\partial \mathcal{J}_1}{\partial Y} &= \frac{\Omega - \Omega^\flat}{2}, \\
    \frac{\partial \mathcal{J}_1}{\partial B} &= \Pi B F_w - \frac{\Omega + \Omega^\flat}{2}(B^\flat)^\dagger.
\end{align}

For the second-order cost term \( \mathcal{J}_2 = \mathrm{Tr}(Q_2 \mathbf{S}_2) \), let \( \Pi_2 \) solve the second-order adjoint Lyapunov equation:
\[
(A \otimes I + I \otimes A)\Pi_2 + \Pi_2(A^\dagger \otimes I + I \otimes A^\dagger) + Q_2 = 0.
\]
Define the operator:
\[
\Omega := \operatorname{Tr}_1(\Pi_2 \mathbf{S}_2) + \operatorname{Tr}_2(\Pi_2 \mathbf{S}_2),
\]
then the Fréchet derivatives of \( \mathcal{J}_2 = \operatorname{Tr}(Q_2 \mathbf{S}_2) \) are given by:

\begin{align}
    \frac{\partial \mathcal{J}_2}{\partial Y} &= \tfrac{1}{2}(\Omega - \Omega^\flat),
\end{align}

\begin{align}
\label{eq:DJ2_dB_final}
\begin{aligned}
    \frac{\partial \mathcal{J}_2}{\partial B} &= 
    \operatorname{Tr}_2 \left( \Pi_2 (I \otimes \mathbf{S}_1) \right) B F_w +
    \operatorname{Tr}_1 \left( \Pi_2 (\mathbf{S}_1 \otimes I) \right) B F_w \\
    &\quad + \operatorname{Tr}_1 \left( M_4 (B^\dagger \otimes I) \right) +
    \operatorname{Tr}_2 \left( M_4 (I \otimes B^\dagger) \right) \\
    &\quad + \operatorname{Tr}_1 \left( M_2 (B \otimes I) \right) +
    \operatorname{Tr}_2 \left( M_3 (I \otimes B) \right) \\
    &\quad - \tfrac{1}{2}(\Omega + \Omega^\flat)(B^\flat)^\dagger.
\end{aligned}
\end{align}

Here, the definition of matrices \( M_1, M_2, M_3, M_4 \) can be found in Appendix~\ref{appendix:ALE}.

These expressions enable the full gradient of the energy cost to be computed with respect to the physical parameters \( Y \) and \( B \), allowing variational optimization via chain-rule propagation to any parameter vector \( \boldsymbol{\theta} \), such as using gradient-based optimizers like Adaptive Moment Estimation (ADAM)~\cite{kingma2014adam}.

\bibliography{apssamp}

\end{document}